\documentclass[notitlepage,article,amsmath,onecolumn]{revtex4-1}
\usepackage[skip=-40pt]{caption}
\usepackage{epsfig,psfrag,graphicx,bm,color}
\usepackage{graphicx}
\graphicspath{  {images/}  }
\usepackage{mathptmx}      % use Times fonts if available on your TeX system
\usepackage{latexsym}
\def\beq{\begin{equation}}
\def\eeq{\end{equation}}
\def\bea{\begin{eqnarray}}

\def\eea{\end{eqnarray}}

\begin{document}

%\title{Generalized geodesic deviation equations and entanglement first law in d-dimensions}
\title{Holographic entanglement first law for $d+1$ dimensional rotating cylindrical black holes}
%\title{The change of holographic entanglement entropy of $d+1$ dimensional rotating stationary cylindrical %black hole and  BTZ black hole in $d+1$ dimensions}

\author{Hamideh Nadi}
\email{h.nadi@ph.iut.ac.ir}
\author{Behrouz Mirza}
\email{b.mirza@cc.iut.ac.ir}
\author{Zeinab Sherkatghanad}
\email{z.sherkat@ph.iut.ac.ir}
\author{Zahra Mirzaiyan}
\email{z.mirzaiyan@ph.iut.ac.ir}

\affiliation{Department of Physics,
Isfahan University of Technology, Isfahan 84156-83111, Iran}

\begin{abstract}
We calculate the holographic entanglement entropy for the rotating cylindrical black holes in $d+1$ dimensions as perturbations over $AdS_{d+1}$. This is accomplished based on the first order variation of the area functional in arbitrary dimensions. For these types of black holes, the angular momentum  appears at the first order of the perturbative expansion of the holographic entanglement entropy for spacetime dimensions of d +1 $\geq$ 4. We obtain a form of holographic entanglement first law in the presence of both energy and angular momentum.

\end{abstract}
\maketitle
%%%%%%%%%%%%%%%%%%%%%%%%%%%%%%%%%%%%%%%%%%%%%%%%%%%%%

%%%%%%%%%%%%%%%%%%%%%%%%%%%%%%%%%%%%%
\section{Introduction}
%%%%%%%%%%%%%%%%%%%%%%%%%%%%%%%%%%%%%%
Entanglement entropy (EE), as a common quantity in quantum field theory (QFT), is investigated in various field including that of quantum information. Assume a given system and dividing it into two parts of $A$ and $B$, the EE for the subsystem $A$ is defined by the von Neumann entropy $S_A=-Tr (\rho_{A} log \rho_{A})$, where $\rho_{A}$  is the reduced density matrix of $A$. $\rho_{A}$ is obtained by tracing over the density matrix for the total system $\rho_{tot}$, with respect to the degrees of freedom of $B$: $\rho_{A}=Tr_B \rho_{tot}$. This entropy is used as a measure of quantum correlations between the subsystems $A$ and $B$. In higher dimensional QFTs, it is difficult to calculate the EE values for an arbitrary subsystem $A$. Ryu and Takayanagi (RT) used the holographic principle to propose a holographic description of this quantity. According to this principle, there is a correspondence between the gravitational theory and the one lower dimensional non-gravitational theory. If the gravitational theory is defined in the Anti-de-Sitter (AdS) space, its dual is a conformal field theory (CFT) located on the boundary of this space. This duality is called the AdS/CFT  correspondence \cite{Maldacena,Aharony}. These dualities may be described in QFT, by considering such  non-local observables as two point correlation functions which are dual to the geodesic length and Wilson loops which are dual to the minimal surface area \cite{D. Galante:2012}. RT proposed a relationship to hold  between the EE of the subsystem $A$ on the $d$ dimensional CFT side and the area of $d-1$ dimensional minimal surfaces anchored at the boundary of $AdS_{d+1}$ on the entangling surface \cite{RT,RT1}. Therefore, the holographic entanglement entropy (HEE) is given by:
\bea
S_A=\frac{Area (\gamma_A)}{4 G_N^{(d+1)} },
\eea
where, $\gamma_{A}$ is the $d-1$ dimensional minimal surface on $t$ constant slice in the bulk and $G_N^{(d+1)}$ is the $d+1$ dimensional Newton constant. It is known that there are usually UV divergences in HEE, and a regularization method is necessary to remove these divergences. Therefore, the HEE for a deformed geometry is described by subtracting the contribution coming from the background AdS spacetime. The complicated form of $\gamma_{A}$ for the metric of $d+1$ dimensional black holes in the bulk often makes the exact calculation of HEE impossible. The HEE for non-rotating BTZ black holes was calculated exactly in Ref. \cite{HRT}. It should be noted that the minimal surface in a non-static spacetime does not live on $t = constant$ slice so that it is impossible to implement the RT proposal. The covariant version of the RT proposal for the bulk metric, which is either non-static or even non-stationary, was presented in \cite{HRT} and the HEE was derived for rotating BTZ black holes as a stationary but non-static spacetime.

Various perturbative methods may be used to evaluate the HEE in the presence of desired non-rotating black holes in the bulk \cite{Ling, Pedraza,1212.1164,Sun}. In this situation, the following relation may be derived for $\Delta S_A$:
\bea
 \Delta S_A =S_{A} -S_{A}^{(0)},
\eea
where, $S_{A}$  is the HEE for an asymptotically AdS metric which is an excitation over pure AdS and $S_{A}^{(0)}$ is the HEE in pure AdS. In Ref. \citep{1212.1164} it is argued that in the asymptotically AdS background, the HEE for the very small boundary subsystem obeys the following relation which is similar to the first law of thermodynamics when the energy of the subsystem, $\Delta E_A$, is raised to the above ground state
 \bea
\Delta S_A=\frac{1}{T_E} \Delta E_A,
\eea
where, $T_E$ is the universal entanglement temperature proportional to the inverse of the subsystem size. The above relation means that the variation in $S_A$ is given by physical observables. Also in Ref. \cite{Alishahiha,Caputa,J. F. Pedraza,Pankaj,Karar,Bhatta,Soltan} entanglement thermodynamics is investigated in various asymptotically AdS background by using the HEE. For example  in Ref. \cite{Alishahiha} the HEE for a very small exited subsystem at the dual CFT is obtained in terms of energy, which is similar to the the first law of thermodynamics. In Ref. \cite{Ghosh}, 2+1 dimensional rotating BTZ black holes are taken as perturbations over the $AdS_{2+1}$ spacetime. Then the HEE for a (2+1) dimensional rotating BTZ black hole was obtained up to the second order of the metric perturbation. For rotating BTZ black holes, energy appeared in the first order of HEE expansion and no angular momentum term contributed. At the second order, however, both energy and angular momentum were observed to contribute. Thus at the second order of perturbation, the first law of entanglement thermodynamics was presented with an extra term due to the presence of angular momentum. So far, no computation have been made for the HEE of higher dimensional rotating black holes. We do this for the first time in this paper.

In this paper, we derive a perturbative expression for the HEE of  $d+1$ dimensional rotating cylindrical black holes. We see that in these black holes ($d>2$), unlike in the rotating BTZ ones, the energy  occurs along with the angular momentum at the first order of perturbation. Thus at the first order the holographic entanglement first law contain both the entanglement temperature and the entanglement angular velocities. 

Our paper is organized as follows. In Section II, we consider a black hole as a perturbation around the $AdS_{d+1}$. This perturbation leads to a deviation of the minimal surface in the bulk. We, therefore, derive an expression for the variations in the area of the $d-1$ dimensional minimal surface up to the first order of the metric perturbation. In Section III, we parametrize the minimal surface to obtain the local basis for $AdS_{d+1}$ spacetime used in the formula in the previous Section. In Section IV, as an  example we consider $d+1$ dimensional planar AdS black holes in the bulk as perturbations over $AdS_{d+1}$ and calculate the HEE for them. In Section V, we calculate the HEE for the $d+1$ dimensional rotating cylindrical black holes as a perturbation around the $AdS_{d+1}$ using our covariant method. We also investigate the entanglement thermodynamics in $d+1$ dimensions for these rotating black holes. In this way, the entanglement temperature and the entanglement angular velocities are obtained as functions of the rotation parameters of rotating cylindrical black holes.
%%%%%%%%%%%%%%%%%%%%%%%%%%%%%%%%%%%%
\section{FIRST ORDER VARIATION OF AREA FUNCTIONAL}
%%%%%%%%%%%%%%%%%%%%%%%%%%%%%%%%%%%
In order to calculate the HEE in $AdS_{d+1}$, it is essential to find the area of the $d-1$  dimensional minimal surface. We assume a black hole as a small perturbation on $AdS_{d+1}$ that leads to the deviation of the minimal surface. The variations in the minimal surface area depend on the changes in the bulk metric and the embedding functions:
\bea
\delta A=A(G+\delta G, x^{\mu}+\delta x^{\mu})-A(G, x^{\mu}).
\eea
We may obtain an expression for the variations in  the minimal surface area and the changes in HEE up to the first order. For a $d+1$ spacetime, the area functional of the minimal surface in $d-1$ dimensions is of the following form
\bea\label{a}
A=\int d^{d-1} \zeta \ \sqrt{h},
\eea
where,
\bea\label{h}
h=det \ h_{ab}\  \ , \ \ h_{ab}=g_{\mu \nu}^{(0)} \partial_a x^{\mu} \partial_b x^{\nu} \ \ , \ \ \partial_a=\frac{\partial}{\partial \zeta ^a},
\eea
in which, $g_{\mu \nu}^{(0)}$ is the metric of $AdS_{d+1}$, $h_{ab}$ is the induced metric on $d-1$  dimensional surface, and $a=1,...,(d-1)$. It is initially assumed that the spacetime is not perturbed so that the geodesic equation for the minimal surface is obtained in this unperturbed spacetime. To get the geodesic equation, the area functional in Eq. (\ref{a}) needs to be varied:
\bea\label{deltaa}
\delta A=\int d^{d-1} \zeta \   \delta \sqrt{h}.
\eea
With some simple calculations we get to the Gauss equation (See appendix A):
\bea\label{geodesic}
\partial_a(\sqrt{h}\ h^{a b} \partial_b x^{\mu}) +\sqrt{h}\ h^{ab} \Gamma^{\mu}_{\nu \rho} \partial_a x^{\nu} \partial_b x^{\rho} =0.
\eea
This is, indeed, the geodesic equation for the initial $d-1$ dimensional minimal surface. We now turn to a small perturbation around $AdS_{d+1}$. We want to derive the difference between the minimal surface areas of the  pure $AdS_{d+1}$ and the perturbation over it. Again, we need to calculate the variation in the area functional (\ref{a}). In this case, $\delta g_{\mu \nu}^{(0)}$ can be written as follows
\bea\label{geodesic}
\delta g_{\mu \nu}^{(0)}=\frac{\partial g_{\mu \nu}^{(0)}}{\partial x^{\rho}}\delta x^{\rho}+g_{\mu \nu}^{(1)},
\eea
where, $g_{\mu \nu}^{(1)}$ is a perturbation over the metric $g_{\mu \nu}^{(0)}$. It is clear that in this case the variation in the area functional includes the geodesic equation and a new term in Eq. (\ref{deltaAf}) that arises from the perturbation in the metric. The variation in the area then becomes
\bea\label{deltaAf}
\delta A=\int d^{d-1} \zeta \ \frac{1}{2} \sqrt{h}\  h^{ab} \ [ g_{\mu \nu}^{(1)} \partial_a x^{\mu} \partial_b x^{\nu}].
\eea
Thus, the variation in  the minimal surface area in the first order of the metric perturbation is derived, which is proportional to the change in the HEE of the pure $AdS_{d+1}$. This method is equivalent to covariant method \cite{HRT} (See appendix B to clarify this issue). Thus, one needs to derive the induced metric, $h^{ab}$, and the local basis, $\partial_a x^{\mu}$, for the pure $AdS_{d+1}$ to compute the modification of HEE in the presence of black holes in the bulk. Also, the metric perturbation $g_{\mu \nu}^{(1)}$ is computed using the Fefferman-Graham expansion close to the boundary in the following calculations for the $d+1$ dimensional planar AdS black holes and the rotating cylindrical black holes in $d+1$ dimensions.
%%%%%%%%%%%%%%%%%%%%%%%%%%
\section{ parametrized minimal surface and solution of equation of motion}
%%%%%%%%%%%%%%%%%%%%%%%%%%%%%%%%%%%%
Let us consider the poincare coordinate where the boundary of $AdS_{d+1}$ is located at $z=0$; we have
\bea\label{b}
ds^2=\frac{1}{z^2}(-dt^2+dz^2+\sum_{i=1}^{d-1} dx_i^2),
\eea
where, $g_{\mu \nu}^{(0)}=\frac{1}{z^2}(-1,1,1,....)$. We consider that the subsystem at the boundary of $AdS_{d+1}$ is a strip with a length of $l$ in a time constant slice and also $-l/2<x_1<l/2$, $-L/2<x_i<L/2$ for $i=2,....d-1$. Considering the black holes as perturbations over $AdS_{d+1}$, it will be necessary to find $\partial_a x^{\mu}$, the local basis of the pure AdS, for use in expression (14). For this purpose, the $(d-1)$ dimensional surface is parametrized with respect to parameters $\tau$ and $\sigma_i$ as $[\zeta^a]=[\tau,\sigma_i]$, while $-\infty<\tau<\infty$ and $-L/2<\sigma_i<L/2$.  Therefore, the area functional in Eq. (\ref{a}) is given by:
\bea\label{NG}
A=\int d^{d-2} \sigma\ d \tau \sqrt{h},
\eea
which is the Nambo-Goto action related to a geodesic surface. Also,  $\sqrt{h}$ in this case plays the role of a Lagrangian. To simplify our calculations, we consider the coordinates in the surface such that $x^\mu$ becomes a function of one of parameters $\sigma_i$ or $\tau$. This assumption within the diagonal form of $AdS_{d+1}$ metric makes a diagonal induced metric; i.e, $h_{\tau \sigma_i}=0$. If we consider $AdS_{d+1}$ in the bulk, the components of the induced metric in Eq. (\ref{h}) can be rewritten in the following form:
\bea\label{htt}
&&h_{\tau \tau}=\frac{1}{z^2} \Big((\frac{\partial x_1}{\partial \tau})^2+(\frac{\partial z}{\partial \tau})^2 \Big),\\\nonumber
&&h_{\sigma_i \sigma_i}=\frac{1}{z^2} (\frac{\partial x_i}{\partial \sigma_i})^2,
\eea
where,
\bea
x_1=x_1(\tau) \  \  , \  \  z=z(\tau) \   \   and  \   \  x_i=x_i(\sigma_i).
\eea
The determinate of the induced metric is given by
\bea\label{hf}
h=(\frac{1}{z^2})^{d-1} \Big((\frac{\partial x_1}{\partial \tau})^2+(\frac{\partial z}{\partial \tau})^2 \Big)\prod_{i=2}^{d-1}(\frac{\partial x_i}{\partial \sigma_i})^2.
\eea
The equation of motion for the Nambo-Goto action in Eq. (\ref{NG}) is
\bea\label{lagrange}
\frac{\partial}{\partial \tau} (\frac{\partial L}{\partial \dot{x}^{\mu}})+\sum_{i=2}^{d-1} \frac{\partial}{\partial \sigma_i} (\frac{\partial L} {\partial x'^{\mu}}) -\frac{\partial L}{\partial x^{\mu}}=0,
\eea
where, $x'^{\mu}=\frac{\partial x^{\mu}}{\partial \sigma_i}$, $\dot{x}^{\mu}=\frac{\partial x^{\mu}}{\partial \tau}$, and  $\mu=0,1,i$. Considering $\sqrt{h}$ as a Lagrangian and solving the above equation for $\mu=0,1$, we get
\bea\label{1}
\frac{\partial z}{\partial \tau}=\pm z^{d-1} \sqrt{1-(\frac{z}{z_{\star}})^{2 (d-1)}},
\eea
\bea\label{2}
\frac{\partial x_1}{\partial \tau}=\frac{z^{2(d-1)}}{z_{\star}^{d-1}},
\eea
 where, $z_{\star}$ is the maximum value of $z$ determined from $\frac{\partial z}{\partial \tau}=0$. One can show that in the limit of $z \to 0$ and for a positive (negative) sign in Eq. (\ref{1}), we have $\tau \to -\infty$ ($\tau \to\infty$). Also, the constant of the integral can be derived at $\tau=0$  while $z=z_{\star}$. The size of the subsystem can be obtained from (21) and (22) in terms of the AdS turning point $z_{\star}$ \cite{RT1}:
\bea\label{l}
l=2 \sqrt{\pi} \ \frac{\Gamma(\frac{d}{2 (d-1)})}{\Gamma(\frac{1}{2 (d-1)})} \  z_{\star}.
\eea
We derived Eqs. (\ref{1}) and (\ref{2}) in the special gauge for $L=\sqrt{h}=1$. The solution of Eq. (\ref{lagrange}) in this gauge for $\mu=i$ is given by:
\bea
\frac{\partial x_i}{\partial \sigma_i}=1  \ \  ,  \ \   i=2,....d-1.
\eea
In what follows, we will investigate the change in the entanglement entropies of both the $d+1$ dimensional planar AdS black hole and the $d+1$ dimensional rotating cylindrical black holes as perturbations over the pure $AdS_{d+1}$ spacetime.
%%%%%%%%%%%%%%%%%%%%%%%%%%%%%%%%%%%%%%%%%%%%%%%%%%%%%%%%%%%%%%%%%%%%%%%%%%%%%%%%%%
\section{holographic entanglement entropy of $d+1$ dimensional planar $AdS$ black holes as perturbation over the pure $AdS_{d+1}$}
%%%%%%%%%%%%%%%%%%%%%%%%%%%%%%%%%%%%%%%%%%%%%%%%%%%%%%%%%%%%%%%%%%%%%%%%%%%%
In this Section, we consider the $d+1$ dimensional planar AdS black holes as perturbations around $AdS_{d+1}$ and determine the changes in the HEE up to the first order. For this purpose, in addition to the local basis of $AdS_{d+1}$ obtained in the previous Section, we need to obtain the metric perturbation $g_{\mu\nu}^{(1)}$ for planar AdS black holes. The metric of the $d+1$ dimensional planar AdS black holes is given by:
\bea\label{metric}
ds^2=-f(r) dt^2+\frac{1}{f(r)} dr^2+r^2 \sum_{i=1}^{d-1} d x_i^2,
\eea
where,
\bea
f(r)=\frac{r^2}{R^2}-r^{2-d} m,
\eea
where, $m$ is the integration constant related to the mass of black hole and $R$ is the radius of the AdS space that we set equal to $1$ throughout our calculations. In general, any spacetime that is asymptotically $AdS_{d+1}$ can be described by Fefferman-Graham coordinates close to the $AdS_{d+1}$ boundary \cite{Fefferman}:
\bea\label{feffer}
ds^2=\frac{dz^2}{z^2}+\frac{(\eta_{\mu \nu}+z^d \gamma_{\mu \nu} +....) dx^{\mu} dx^{\nu}}{z^2}.
\eea
Using the coordinate transformation $\frac{d z}{z}=\frac{d r}{\sqrt{f(r)}}$, we obtain
\bea
 r=\frac{1}{z} (1+\frac{1}{4} m \ z^d)^ {2/d}.
\eea
Then, the metric (\ref{metric}) can be represented in Fefferman-Graham expansion near the boundary $(z=0)$:
\bea
ds^2&=&\frac{dz^2}{z^2}-(\frac{1}{z^2}- g_{tt}^{(1)}+...) dt^2+(\frac{1}{z^2}+g_{x_i x_i}^{(1)}+...)\sum_{i=1}^{d-1}d x_i^2\\\nonumber
&=&\frac{dz^2}{z^2}-\frac{1}{z^2}(1- mz^d (\frac{d-1}{d})+...) dt^2+\frac{1}{z^2} (1+\frac{1}{d}m z^d+...)\sum_{i=1}^{d-1}d x_i^2.
\eea
So, the metric of the planar AdS black holes transformed into (\ref{feffer}), where the non-zero values of $g_{\mu \nu}^{(1)}=z^{d-2} \gamma_{\mu \nu}$ in $d+1$ dimensions is given by
\bea\label{tt}
g_{tt}^{(1)} &=& m z^{d-2} (\frac{d-1}{d}),
\eea
\bea
g_{x_i x_i}^{(1)} &=& \frac{1}{d}m z^{d-2} \ \ \ \ \  for \   \   i=1 .... d-1.
\eea
Substituting the above components, the local basis relations for $AdS_{d+1}$, and Eq. (\ref{htt}) into (\ref{deltaAf}) will yield the perturbation in the entanglement entropy up to the first order:
\bea\label{entropy}
\delta S_1=\frac{\delta A}{4 G_N^{(d+1)}}=\frac{m \ L^{d-2} l^2}{32 (d+1) G_N^{(d+1)}} \frac{\Gamma(\frac{1}{2 (d-1)})^2 \Gamma(\frac{1}{d-1})}{\sqrt{\pi} \Gamma(\frac{1}{2}+\frac{1}{d-1}) \Gamma(\frac{d}{2 (d-1)})^2},
\eea
in which, the integral (\ref{deltaAf}) is evaluated based on $z=0$ to $z=z_{\star}$. Moreover, $z_{\star}$ is substituted with respect to $l$ from Eq. (\ref{l}). So, the modification of HEE up to the first order is proportional to the  black hole mass and the subsystem size. In Ref. \cite{1212.1164}, HEE is calculated for the asymptotically $AdS_{d+1}$ background by using a perturbative method in the very small $l$ limit such that $m l^d<<1$. If we use the method of Ref. \cite{1212.1164} to obtain $\delta S_1$ for the metric (\ref{metric}), we find the same expression (\ref{entropy}) for the HEE of the planar AdS black holes.
%%%%%%%%%%%%%%%%%%
\subsection{Entanglement first law for $d+1$ dimensional planar $AdS$ black holes}
%%%%%%%%%%%%%%%%%%%
According to the AdS/CFT dictionary, the transition from the ground state to an excited state for a system in the dual CFT implies a change from the pure AdS to an asymptotically AdS in the bulk. In thermodynamics, changes in the total energy, $E$, of a system depend on changes in  entropy, $S$. This leads to the first law of thermodynamics and the definition of temperature: $dE=TdS$. It has been argued in Ref. \cite{1212.1164} that the relationship between the entanglement entropy of a very small subsystem in the boundary CFT and the energy of this subsystem is interpreted in terms of the first law of thermodynamics. This relationship was expressed by an effective (entanglement) temperature, $T_{ent}$, that is proportional to the inverse of the subsystem size: $T_{ent}=\frac{\Delta E}{\Delta S}$. As already mentioned, we can write any asymptotically AdS spacetime in the Fefferman-Graham coordinates as in Eq. (\ref{feffer}). In this situation, the observable parameters, like energy and angular momentum, appear as fluctuations about the pure AdS spacetime. The expectation value of the boundary energy momentum tensor is defined by the first perturbative term in the metric as follows \cite{Solodukhin, Myers, 9902121}:
\bea\label{energytensor}
<T_{\mu \nu}>=\frac{d}{16 \pi G} \gamma_{\mu \nu}.
\eea
An increased amount of energy for an excited system is given by the first component of this tensor:
\bea
\Delta E=\int d^{d-1} x \ <T_{tt}>.
\eea
Using the above equations and the relation (\ref{tt}) for the $d+1$ dimensional planar AdS black holes, we have
\bea
\Delta E=\int d^{d-1} x \frac{d-1}{16 \pi G} m=\frac{d-1}{16 \pi G} m \ l \ L^{d-2}.
\eea
From the above equation and Eq. (\ref{entropy}), like what is said in Ref. \cite{1212.1164} we get the following ratio between energy and the HEE of the subsystem and define it the entanglement temperature:
\bea
T_E=\frac{2 (d+1)(d-1)}{\sqrt{\pi}} \frac{\Gamma(\frac{1}{2}+\frac{1}{d-1}) \Gamma(\frac{d}{2 (d-1)})^2}{\Gamma(\frac{1}{d-1}) \Gamma(\frac{1}{2 (d-1)})^2} l^{-1}.
\eea
It is thus seen that the entanglement temperature depends only on the shape of the subsystem and changes with the inverse of the subsystem size $l$. Thus, the entanglement temperature for a very small excited subsystem is obtained such that the relationship between the entanglement entropy and the increased energy is similar to the first law of thermodynamics.
\section{holographic entanglement entropy of $d+1$ dimensional rotating cylindrical black holes}
%%%%%%%%%%%%%%%%%%%%%%%%%%%%%%%%%%%%
In this Section, which is actually the main part of the article, we obtain the HEE for the $d+1$ dimensional rotating cylindrical  black holes. As already mentioned in the introduction, the typical RT method is not, however, applicable to rotating black holes. Hence, the method proposed in Section II is used to obtain the HEE for $d+1$ dimensional rotating cylindrical black holes. The exact solution for rotating  black holes with cylindrical symmetry in (3+1) dimensions and a negative cosmological constant may be found in \cite{ Lemo}. The generalization of this solution to higher dimensions was presented in \cite{0209238,1702.02416}. The metric is given by:
\bea\label{metricSR}
ds^2&=&-F(r) \Big(\Xi dt-\sum_{i=1}^n a_i \ d\phi_i \Big)^2+r^2 \sum_{i=1}^n (a_i dt-\Xi \ d\phi_i)^2\\\nonumber
&+&\frac{dr^2}{F(r)}-r^2 \sum_{i<j}^n (a_i \ d\phi_j-a_j \ d\phi_i)^2+r^2 \sum_{i=1}^{d-1-n} dx_i^2,
\eea
where, $n =[d/2]$ is the number of rotation parameters $a_i$ and $\Xi=\sqrt{1+\sum_i ^n a_i^2}$. The metric function is
\bea
F(r)=r^2 (1-(\frac{r_+}{r})^d),
\eea
where, $r_+$ is the place of event horizon. We can consider the above metric as perturbation over pure $AdS_{d+1}$ spacetime to calculate the regularized HEE for rotating cylindrical black holes in higher dimensions. In order to obtain the metric fluctuations over AdS in $d+1$ dimensions, $g^{(1)}_{\mu \nu}$, we need to find the Fefferman-Graham expansion close to the boundary. Adopting the definition of the new coordinate, $\frac{d z}{z}=\frac{d r}{\sqrt{F(r)}}$, we can obtain $r$ as follows
\bea
r=\frac{1}{z} (1+\frac{1}{4} r_+^d z^d)^ {2/d}.
\eea
Rewriting the metric in Eq. (\ref{metricSR}) in terms of the new coordinate $z$ near the boundary $(z=0)$, we will have the following metric  in poincare coordinates ($\phi_i \rightarrow \frac{\tilde{x_i}}{l_{AdS}}$):
\bea \label{fff}
ds^2&=&\frac{dz^2}{z^2}-\frac{1}{z^2}(1- r_+^d z^d (\frac{d-1}{d})+...) (\Xi dt-\sum_{i=1}^n a_i d\tilde{x_i})^2\\\nonumber
&+&\frac{1}{z^2} (1+\frac{1}{d} r_+^d z^d+...)\sum_{i=1}^{n} (a_i dt -\Xi d\tilde{x_i})^2\\\nonumber
&-&\frac{1}{z^2} (1+\frac{r_+^d z^d}{d}+...) \sum_{i<j}^n (a_i d\tilde{x_j}-a_j d\tilde{x_i})^2\\\nonumber
&+&\frac{1}{z^2} (1+\frac{r_+^d z^d}{d}+...) \sum_{i=1}^{d-1-n} dx_i^2.
\eea
Thus, the non-zero values of the metric perturbation  coefficients $g_{\mu \nu}^{(1)}$ in $d+1$ dimensions are given by
\bea\label{metriccof}
g_{tt}^{(1)} &=& r_+^d z^{d-2} (\Xi ^2-\frac{1}{d}), \\
g_{t \tilde{x_i}}^{(1)} &=&- a_i \Xi r_+^d z^{d-2}, \\
g_{\tilde{x_i} \tilde{x_i}}^{(1)} &=& r_+^d z^{d-2} ( a_i^2 +\frac{1}{d}),\\
g_{x_i x_i}^{(1)} &=&\frac{1}{d}r_+^d z^{d-2}.
\eea
We assume that $\tilde{x_1}$ and $z$ are functions of $\tau$, and further that $\tilde{x_i}=\tilde{x_i}(\sigma_i)$ for $i=2... n$ and $x_i=x_i(\sigma_i)$ for $i=1...  d-n-1$. Thus, using Eq. (\ref{deltaAf}), the local basis relations for $AdS_{d+1}$ spacetime, Eq. (\ref{htt}), and the above relations for the components of the metric perturbation, the regularized form of HEE up to the first order will be as in (45):
\bea\label{entropy2}
\delta S_1&=&\frac{\delta A}{4 G_N^{(d+1)}}=\frac{r_+^d    L^{d-2}}{4 G_N^{(d+1)}} (\frac{d-1}{2 (d+1)}+\frac{a_1^2}{d+1}+ \frac{(n-1)}{2} \sum_{i=2}^{n} a_i^2)\\\nonumber
&\times &\Big[\frac{ l^2 \Gamma(\frac{1}{2 (d-1)})^2   \Gamma(\frac{1}{d-1})}{4 \sqrt{\pi} (d-1) \Gamma(\frac{d}{2 (d-1)})^2 \Gamma(\frac{1}{2}+\frac{1}{d-1})}\Big],
\eea
in which the integral (\ref{deltaAf}) is evaluated in the interval $0<z<z_{\star}$, $-l/2<\tilde{x_1}<l/2$ and $-L/2<(\tilde{x_i}$ and $ x_i)<L/2$. Moreover, $z_{\star}$ is replaced with respect to $l$ from Eq. (23). It is observed that the HEE for these types of black holes is not only proportional to event horizon but also to the rotating parameters, $(a_1, a_i)$.
%%%%%%%%%%%%%%%%%%
\subsection{Entanglement first law for $d+1$ dimensional rotating cylindrical black holes }
%%%%%%%%%%%%%%%%%%%
rWhen we have a subregion $A$ in QFT, a small change in the density matrix of a pure state $\rho=\rho^{0}+\delta \rho$ makes a small change in the entanglement entropy of $A$, $\delta S_{A}$. As calculated in \cite{Wong}, $\delta S_{A}$ satisfies a local relation similar to the first law of thermodynamics
\bea
\delta S_{A}(x)=\beta_{0} \delta E_{A},
\eea
where, $\delta E_{A}=\delta <T_{00}> Vol(A)$ is the increased energy inside $A$ and $\beta_{0}$ is obtained from the local inverse
entanglement temperature $\beta (x)$ by the relation: $\beta_{0}=\frac{\int_{A} \beta (x)}{Vol(A)}$. Also a generalization of the above relation in the presence of conserved charges $Q_{a}$ and chemical potentials $\mu_{a}$ was obtained in \cite{Wong}. In this case for a state at finite temperature $T$, the density matrix as follows
\bea
\rho=\frac{exp(-\frac{(H-\mu_{a} Q_{a})}{T})}{Z}.
\eea
Thus, the first law becomes general to the following form
\bea\label{s}
\delta S_{A}(x)=\beta_{0}( \delta E_{A}-\mu_{a} \delta Q_{aA}),
\eea
where, $\delta Q_{aA}=\delta <Q_{a}> Vol(A)$. The corresponding density matrix for rotating cylindrical black holes background is written as
\bea
\rho=\frac{exp(-\frac{(H-\Omega_{i} J_{i})}{T})}{Z},
\eea
where, $H$ is the hamiltonian and $\Omega_{i}$ are the angular velocities as the conjugate of the angular momentums $J_{i}$. So the first law that is similar to the Eq. (\ref{s}) can be written by replacing $\mu$ with $\Omega$ and the conserved charge $Q$ with the conserved angular momentum $J$. 

Here we write the entanglement first law for the $d+1$ dimensional rotating cylindrical black holes by using the HEE. According to Ref. \cite{1212.1164,Ghosh}, we define the entanglement temperature and the entanglement angular velocity to satisfy the entanglement first law. The observable parameters can be derived from the expectation value of the boundary energy-momentum tensor in asymptotically AdS spacetimes \cite{9902121}. The presence of the black hole in the bulk corresponds to exciting the system at the boundary. From Eq. (\ref{energytensor}) and the components of the metric perturbation in Eqs. (\ref{metriccof}) and (42), amount of increased energy and angular momentum for the entangling excited subsystem are
\bea\label{E}
\Delta E=\int d^{d-1} x \ <T_{tt}>=\frac{l}{16 \pi G}  L^{d-2} r_+^d (d \ \Xi^2 -1),
\eea
\bea\label{J}
\Delta J_i=-\int d^{d-1} x \  <T_{t\phi_i}>=\frac{l \ d}{16 \pi G}  \ L^{d-2} \ r_+^d  a_i \ \Xi.
\eea
Relation (\ref{entropy2}) for $\delta S_{1}$ can be written in terms of $\Delta E$ and $\Delta J_i$ using Eqs. (\ref{E}) and (\ref{J}). For example, for spacetime dimensions: $d+1=3$ (rotating BTZ black hole), we will have:
\bea
\delta S_1=\frac{l \Delta E \pi}{3},
\eea
that depends only on energy and there is no contribute from angular momentum. Also, for spacetime dimensions: $d+1=4$, we will get:
\bea
\delta S_1=\frac{l \ \Big(3 \Delta E+\sqrt{(- 8 \Delta J_1^2+9 \Delta E^2)}\ \Big)\pi \ \Gamma \left(\frac{1}{4}\right)^2 }{96 \ \Gamma \left(\frac{3}{4}\right)^2}.
\eea
The angular momentum in this dimension is seen to appear as an effective quantity along with the energy. The same holds true for higher dimensions. Here we can express the entanglement first law for rotating cylindrical black holes in the presence of both energy and angular momentum
\bea\label{fl}
d(\Delta S)=\frac{1}{T_E} d (\Delta E)-\sum_{i=1} ^n \frac{\Omega_{iE}}{T_E} d (\Delta J_i),
\eea
where, $\Delta S=\delta S_1$ is the difference in entanglement entropy between the excited state and the ground state or the AdS spacetime up to the first order. Note that we can compute the pressure components along the $\tilde{x_i}$ and $x_i$ directions by using the energy-momentum tensor of the CFT as follows:
\bea
\Delta P|_{{\tilde{x_i} \tilde{x_i}}}=\langle T_{\tilde{x_i} \tilde{x_i}} \rangle=\frac{d r_+^d (a_i^2+\frac{1}{d})}{16 \pi G},\ \  \ \ \Delta P|_{{x_i}{x_i}}=\langle T_{{x_i}{x_i}} \rangle=\frac{ r_+^d}{16 \pi G}.
\eea
It is clear that these two quantities are also functions  of $\Delta E$ and $\Delta J_i$ and have no effect on the first law and in the entanglement first law $V \Delta P$ dose not appear. However, $V \Delta P$ is an  essential term in the first law when  there is another variable in addition to $r_+^d$ and $a_i$  in the metric. For example this is the case for the boosted black brane \cite{Mishra} and some other backgrounds.

When we fix the size of the subsystem ($l$), the entanglement temperature and  angular velocity can be obtained for $d+1$ dimensional rotating cylindrical black holes to satisfy the entanglement first law (\ref{fl}). For example, in spacetime dimensions: $d+1=4$, we have one rotation parameter $(n=1)$. It is, therefore, clear  from Eqs. (\ref{entropy2}), (\ref{E}) and (\ref{J}) that: $\delta S_1=\delta S_1 (a_1,r_+)$, $\Delta E=\Delta E (a_1,r_+)$ and $\Delta J_1=\Delta J_1 (a_1,r_+)$. Now we can use the method of \cite{Hosseini} to calculate the partial derivatives (See appendix C for an introduction to the bracket notation).\\
\bea\label{p}
\frac{\partial(\delta S_1)}{\partial(\Delta E)}|_ {\Delta J_1 \ fixed}&=&\frac{\lbrace \delta S_1,\Delta J_1\rbrace_{a_1,r_+}}{\lbrace \Delta E,\Delta J_1\rbrace_{a_1,r_+}},\\\nonumber
\frac{\partial(\delta S_1)}{\partial(\Delta J_1)}|_ {\Delta E \ fixed}&=&\frac{\lbrace \delta S_1,\Delta E\rbrace_{a_1,r_+}}{\lbrace \Delta J_1,\Delta E\rbrace_{a_1,r_+}}.
\eea
So the entanglement temperature and  angular velocity are given by
\bea
\frac{1}{T_E}&=&\frac{\partial(\delta S_1)}{\partial(\Delta E)}|_ {\Delta J_1 \ fixed}=\frac{l \ \pi \ \Big(1+a_1^2 \Big)\Gamma \left(\frac{1}{4}\right)^2}{8\ \Big(2+a_1^2 \Big)\Gamma \left(\frac{3}{4}\right)^2},\\\nonumber
\frac{\Omega_{1E}}{T_E}&=&-\frac{\partial(\delta S_1)}{\partial(\Delta J_1)}|_ {\Delta E \ fixed}=\frac{l \ \pi \ \Big(a_1 \sqrt{1+a_1^2} \Big)\Gamma \left(\frac{1}{4}\right)^2}{12\ \Big(2+a_1^2 \Big)\Gamma \left(\frac{3}{4}\right)^2}.\\\nonumber
\eea
For $d+1=5$, there are two rotation parameters $(n=2)$; so, $\delta S_1=\delta S_1 (a_1,a_2,r_+)$, $\Delta E=\Delta E (a_1,a_2,r_+)$, $\Delta J_1=\Delta J_1 (a_1,a_2,r_+)$, and $\Delta J_2=\Delta J_2 (a_1,a_2,r_+)$. Using the general form of Relation (\ref{p}) for the functions of the three variables, we obtain the partial derivative and, thereby, the entanglement temperature and angular velocity as follows:\\
\bea
\frac{1}{T_E}&=&\frac{\partial(\delta S_1)}{\partial(\Delta E)}|_ {\Delta J_1,\Delta J_2 \ fixed}=\frac{l\ \sqrt{\pi}\Big(3+4 a_1^2+a_2^2\Big)\Gamma \left(\frac{1}{6}\right)^2\ \Gamma \left(\frac{1}{3}\right)}{30\ \Big(3+2(a_1^2+a_2^2)\Big)\Gamma \left(\frac{2}{3}\right)^2\ \Gamma \left(\frac{5}{6}\right)},\\\nonumber
\frac{\Omega_{1E}}{T_E}&=&-\frac{\partial(\delta S_1)}{\partial(\Delta J_1)}|_{\Delta E,\Delta J_2 \ fixed}=\frac{l\ \sqrt{\pi}\ a_1\Big(2+2 a_1^2+ a_2^2\Big)\Gamma \left(\frac{1}{6}\right)^2\ \Gamma \left(\frac{1}{3}\right)}{20\ \Xi \Big(3+2(a_1^2+a_2^2)\Big)\Gamma \left(\frac{2}{3}\right)^2\ \Gamma \left(\frac{5}{6}\right)},\\\nonumber
\frac{\Omega_{2E}}{T_E}&=&-\frac{\partial(\delta S_1)}{\partial(\Delta J_2)}|_{\Delta E,\Delta J_1 \ fixed}
=-\frac{l\ \sqrt{\pi}\ a_2\Big(1+a_2^2\Big)\Gamma \left(\frac{1}{6}\right)^2\ \Gamma \left(\frac{1}{3}\right)}{20\ \Xi \Big(3+2(a_1^2+a_2^2)\Big)\Gamma \left(\frac{2}{3}\right)^2\ \Gamma \left(\frac{5}{6}\right)}.
\eea
We can use the general form of the partial derivative formula for the functions of $(n+1)$ variables. Employing this procedure for higher dimensions, we can obtain the entanglement temperature in arbitrary dimensions:
\bea\label{TE}
\frac{1}{T_E}&=&\frac{\Gamma \left(\frac{1}{2(d-1)}\right)^2\ \Gamma \left(\frac{1}{d-1}\right)}{\Gamma \left(\frac{d}{2(d-1)}\right)^2\ \Gamma \left(\frac{1}{d-1}+\frac{1}{2}\right)}\\\nonumber
&\times&\frac{l\ \sqrt{\pi}\ \Big((d-1)+2(d-2)a_1^2+((3d-1)-(d+1)n) \sum_{i=2}^n a_i^2\Big)}{2(d^2-1)\ \Big((d-1)+(d-2) \sum_{i=1}^n a_i^2 \Big)}.
\eea
Also, the entanglement angular velocities, $\Omega_{1E}$ as the conjugate of $\Delta J_1$, and  $\Omega_{jE}$ as the conjugate of $\Delta J_j$ in arbitrary dimensions are given by
\bea\label{OT}
\Omega _{1_E}&=&\frac{2\ a_1}{d\ \Xi}\times \Big(\frac{1}{(d-1)+2(d-2)a_1^2+((3d-1)-(d+1)n)\sum_{i=2}^n a_i^2}\Big)\\\nonumber
&\times&\Big((d-1)(d-2)(1+a_1^2)+((d(d-2)+5)-(d+1)n)\sum_{i=2}^n a_i^2\Big),
\eea
\bea
\Omega _{j_E}&=& \frac{-2\ a_j}{d\ \Xi}\times \Big(\frac{1}{(d-1)+2(d-2)a_1^2+((3d-1)-(d+1)n)\sum_{i=2}^n a_i^2}\Big)\\\nonumber
&\times& \Big(a_1^2\ (d(d-1)-2)(n-2)+(d-1)(1+\sum_{i=2}^n a_i^2)(-(2d+1)+(d+1)n) \Big),
\eea
that is, $j=2,3,... n$. Using Relations (\ref{TE}) and (\ref{OT}) when $d=2\ (n=1)$, we have $T_E=\frac{3}{\pi \ l}$ and $\Omega_{1E}=0$, which is consistent with the result reported in \cite{Ghosh} for a rotating BTZ black hole. For $d=2$, therefore, the entanglement temperature changes with the inverse of $l$ and the entanglement angular velocity vanishes at the first order. This means that for the rotating BTZ black hole the leading order of the entanglement angular velocity appears at the second order of perturbation. It may also be noted  from the above relations that for $d=3,4,5,....$ the entanglement temperature depends on both $l$ and rotation parameters $a_i$ and it is interesting that the entanglement angular velocity appears in the first order and depends on rotation parameters as well. Dependence of the entanglement temperature on the subsystem size and rotation parameters is depicted in Figs.1 and 2 for $d=3$,$4$, respectively. Also, the entanglement angular velocities are plotted as functions of rotation parameters in Figs.1 and 2 for $d=3$,$4$, respectively.
\begin{figure*}
	\centering
	\begin{tabular}{ccc}
		\rotatebox{0}{
			\includegraphics[width=0.3\textwidth,height=0.42\textheight ]{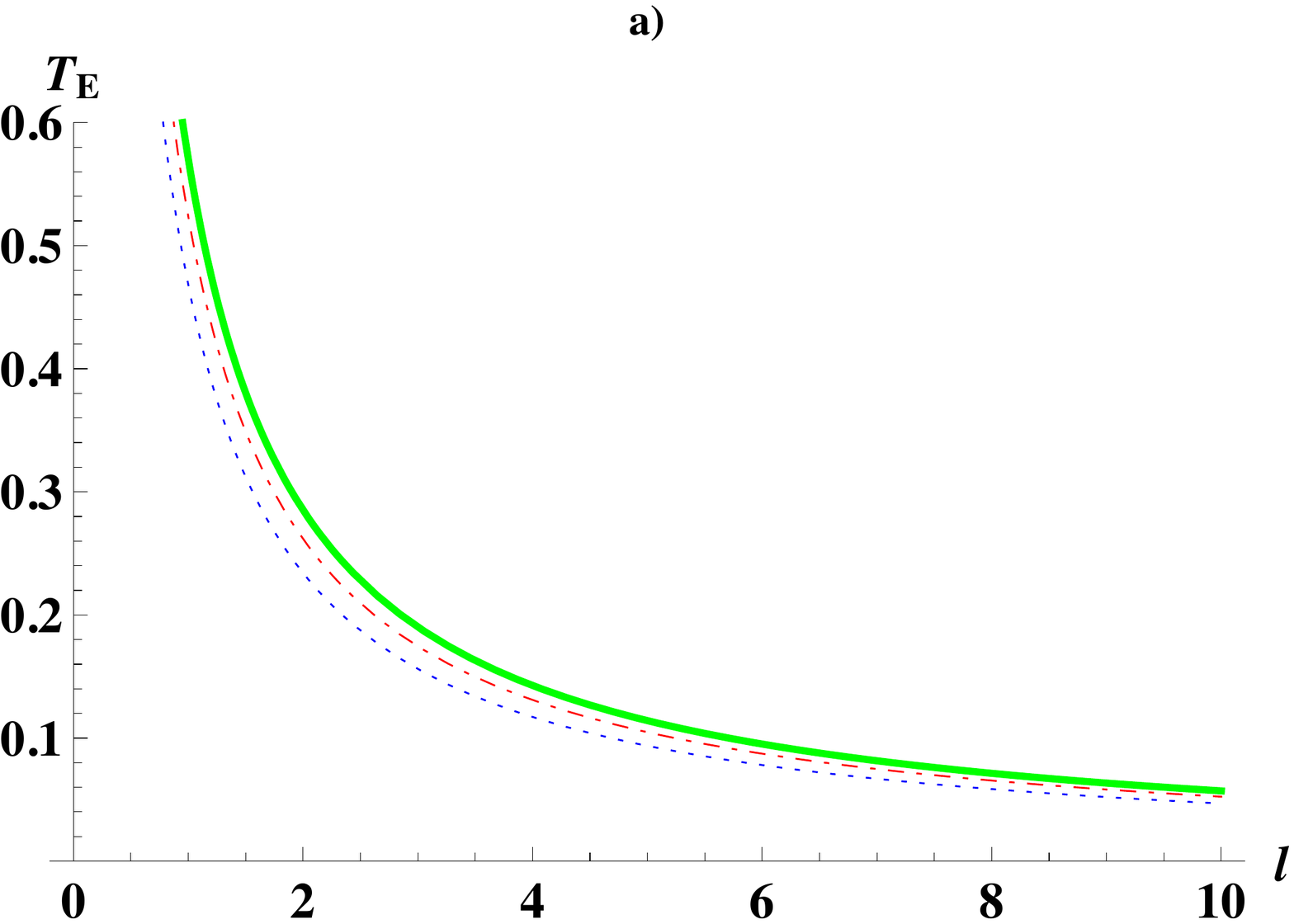}}&
		\rotatebox{0}{
			\includegraphics[width=0.3\textwidth,height=0.42\textheight]{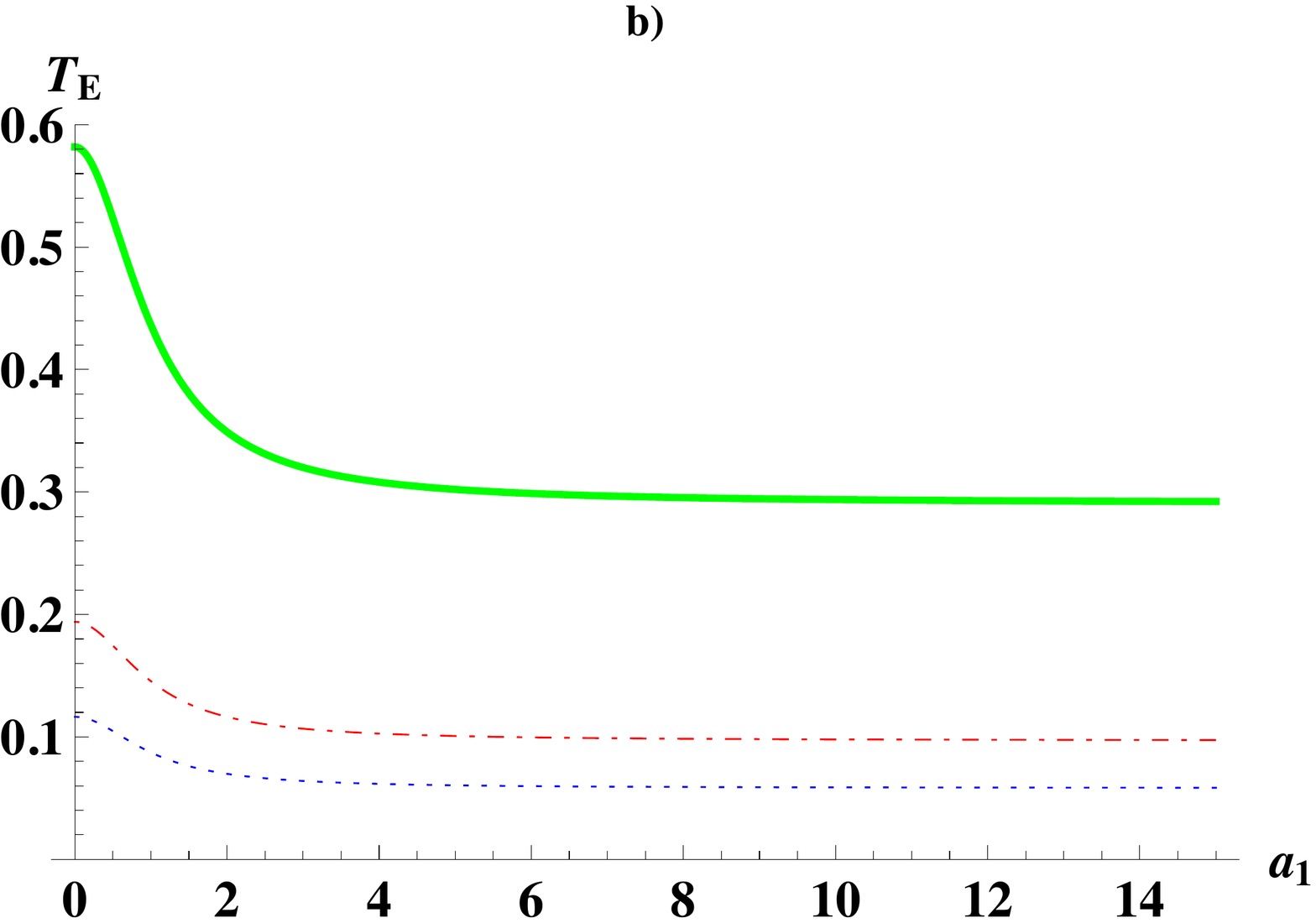}}&
		\rotatebox{0}{
			\includegraphics[width=0.3\textwidth,height=0.42\textheight]{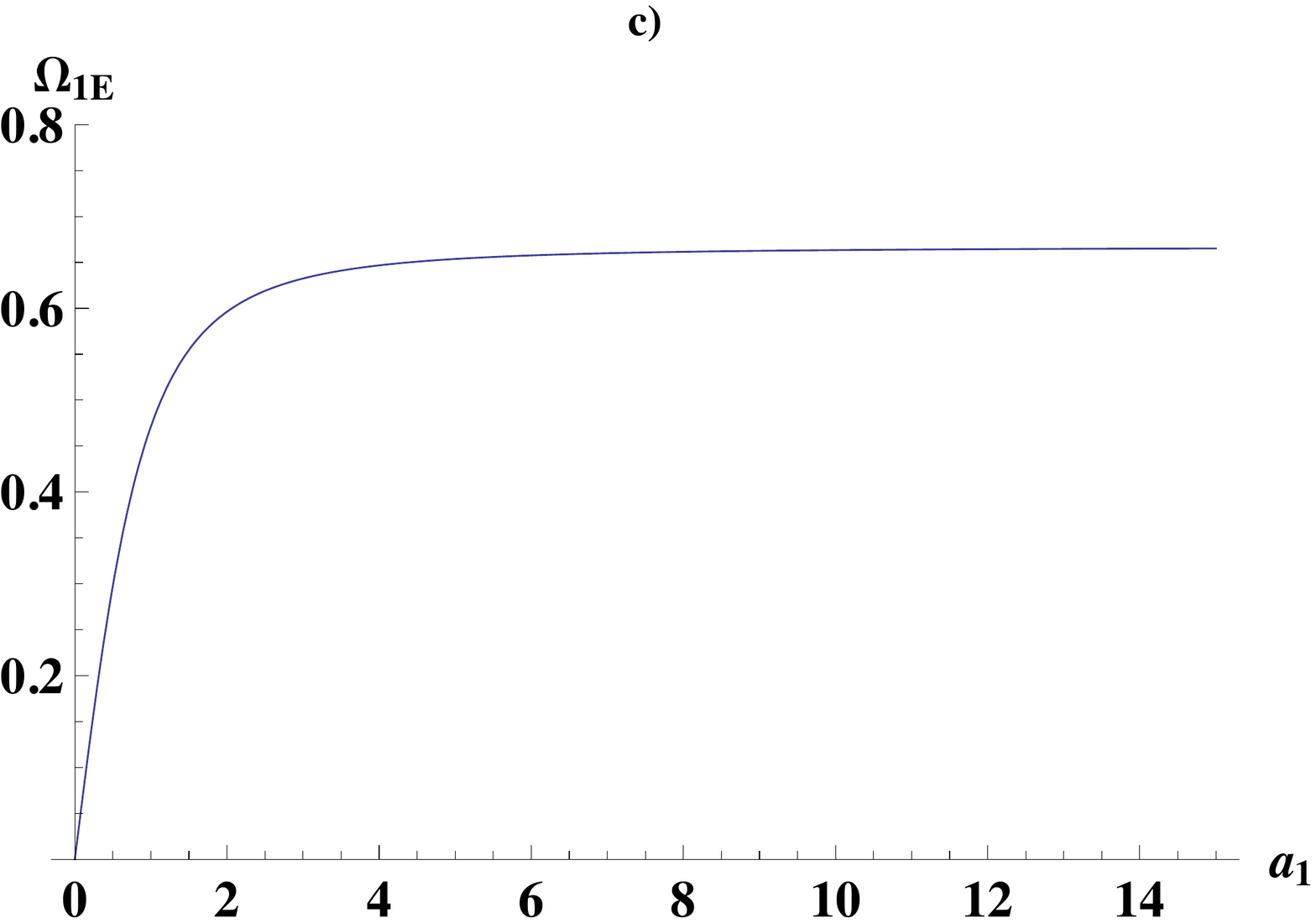}}\\
		\\
		%width=0.4\textwidth,height=0.33\textheight
	\end{tabular}
	\caption{ \it{Plot a,  entanglement temperature $T_{E}$ as a function of subsystem size $l$ for spacetime dimensions: $d+1=4$  for $a_1=0.2$ (green thick line), $a_1=0.5$ (red dot-dashed line), and $a_1=0.8$ (blue dot line); Plot b, $T_{E}$ as a function of rotation parameter $a_1$ for spacetime dimensions: $d+1=4$ for $l=1$ (green thick line), $l=3$ (red dot-dashed line), and $l=5$ (blue dot line); and Plot c,  entanglement angular velocity $\Omega_{1E}$ as a function of $a_1$ for spacetime dimensions: $d+1=4$.}
	}\label{figure:TL}
\end{figure*}
\begin{figure*}
	\centering
	\begin{tabular}{cc}\vspace{-3.5 cm}
		\rotatebox{0}{
			\includegraphics[width=0.4\textwidth,height=0.42\textheight]{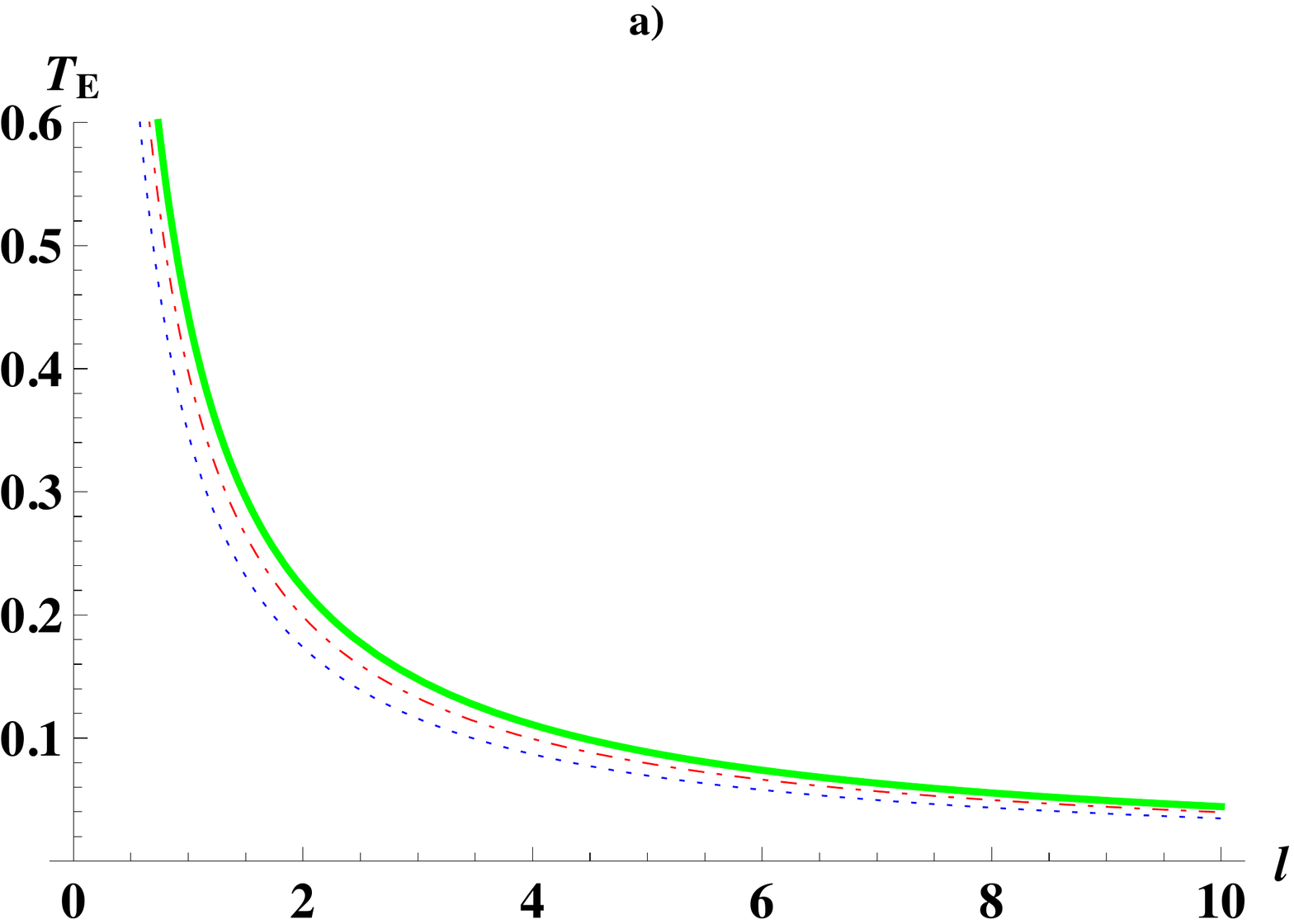}}&
		\rotatebox{0}{
			\includegraphics[width=0.4\textwidth,height=0.42\textheight]{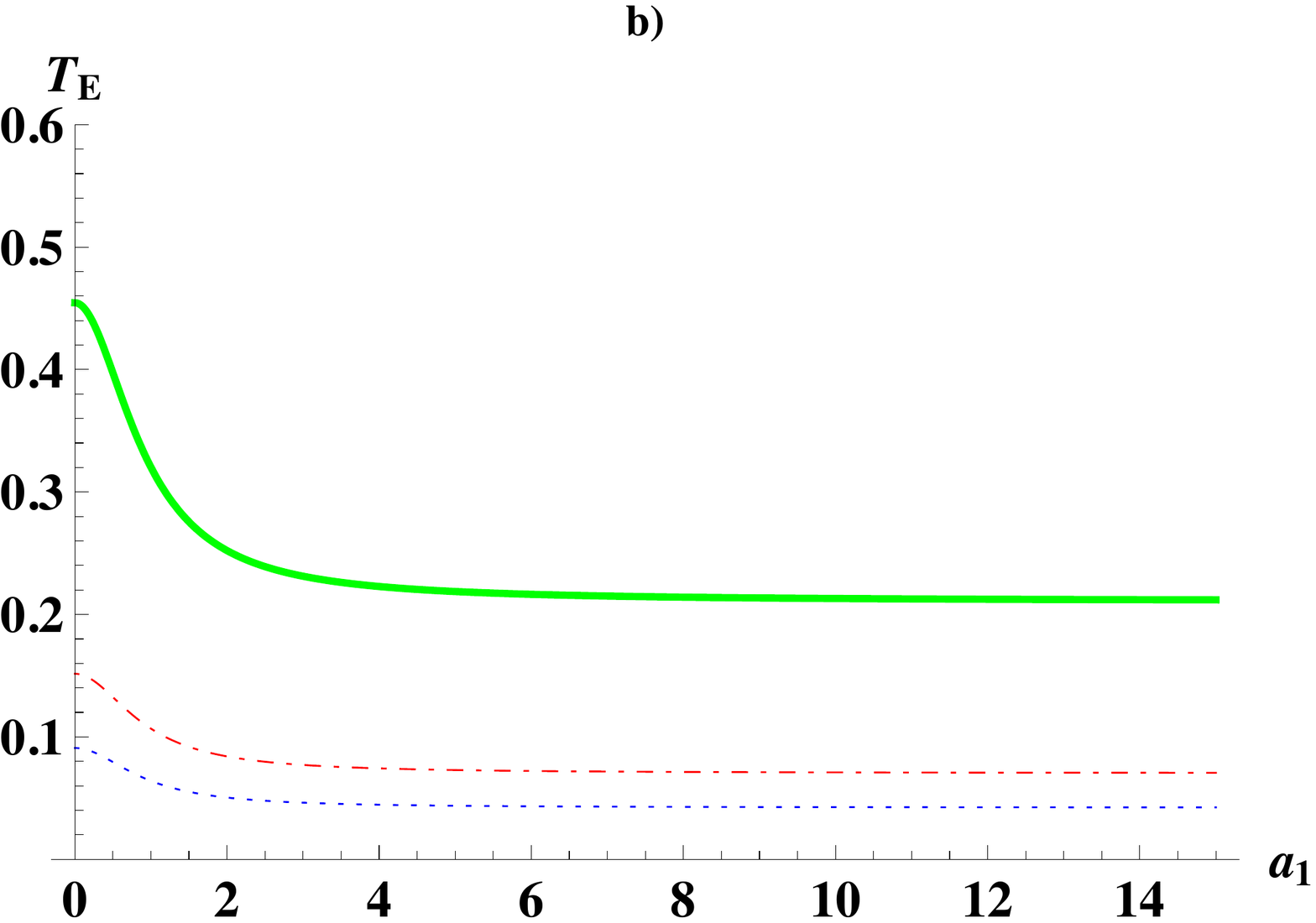}}\\ 
		\rotatebox{0}{
			\includegraphics[width=0.4\textwidth,height=0.42\textheight]{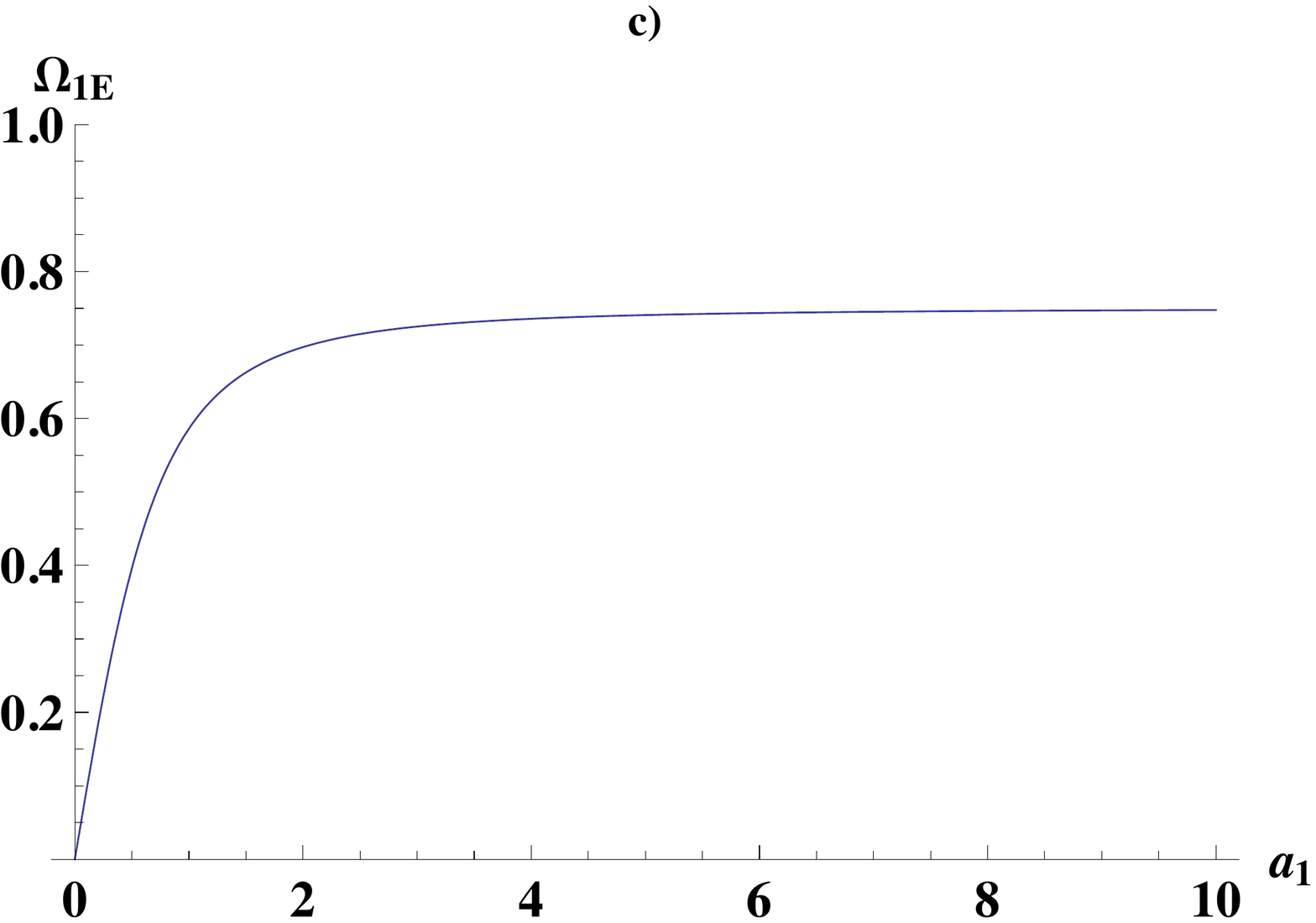}}&
		\rotatebox{0}{
			\includegraphics[width=0.4\textwidth,height=0.42\textheight]{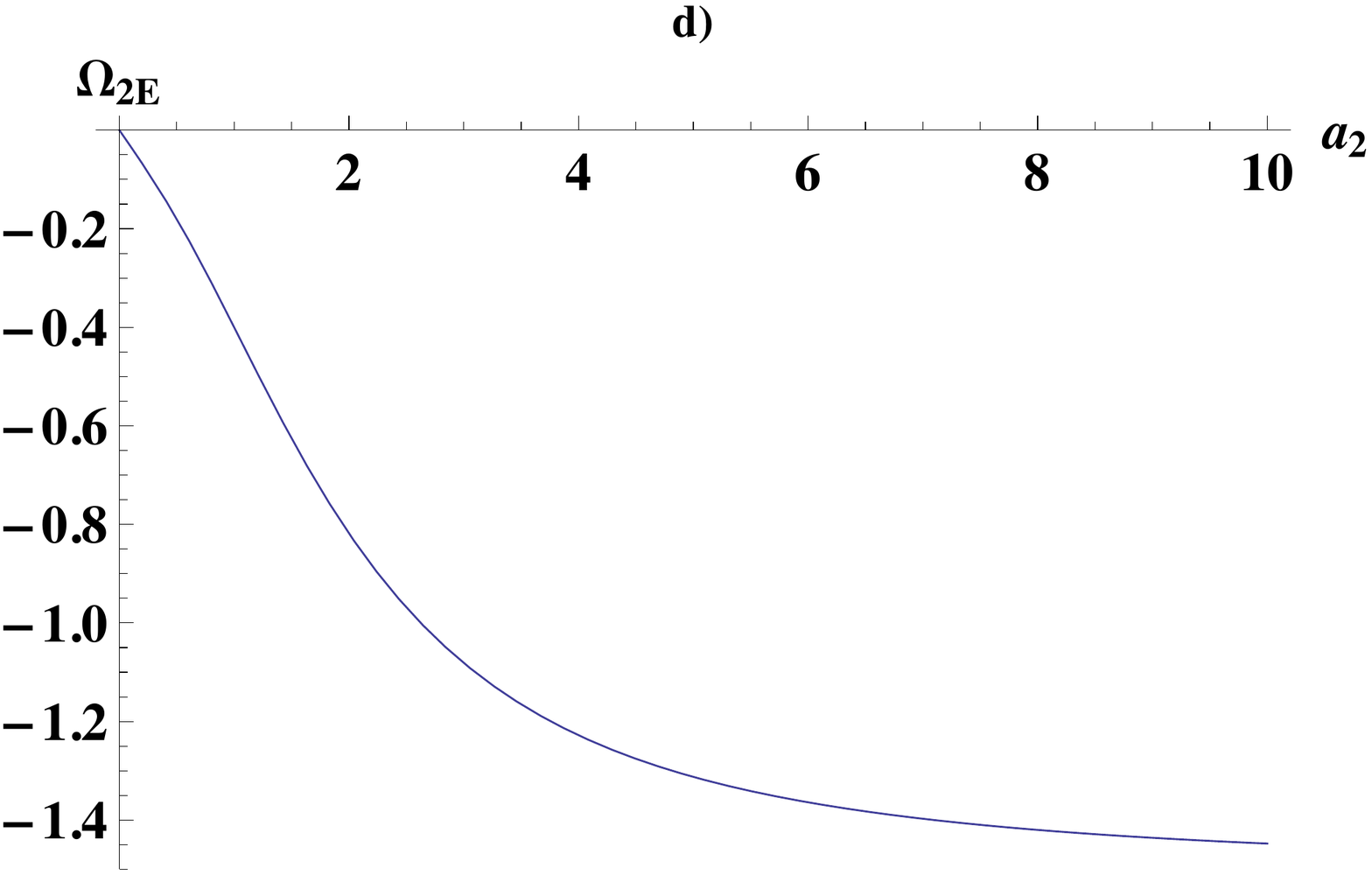}}\\
		\\
		%width=0.4\textwidth,height=0.33\textheight
	\end{tabular}
	\caption{ \it{Plot a,  entanglement temperature $T_{E}$ as a function of subsystem size $l$ for spacetime dimensions: $d+1=5$ and $a_2=0.5$ for $a_1=0.2$ (green thick line), $a_1=0.5$ ( red dot-dashed line), and $a_1=0.8$ (blue dot line); Plot b, $T_{E}$ as a function of $a_1$ for spacetime dimensions: $d+1=5$ and $a_2=0.5$ for $l=1$ (green thick line), $l=3$ (red dot-dashed line), and $l=5$ (blue dot line); Plot c,  entanglement angular velocity $\Omega_{1E}$ as a function of $a_1$ for spacetime dimensions: $d+1=5$ and $a_2=0.5$; and Plot d, $\Omega_{2E}$ as a function of $a_2$ for spacetime dimensions: $d+1=5$ and $a_1=0.5$.}
	}\label{figure:TL}
\end{figure*}

%%%%%%%%%%%%%%%%%%%%%%%%
\section{Conclusion}
%%%%%%%%%%%%%%%%%%%%%%%%%%%%%%%%%%%%%%%%%%%%%%%%
In this paper, we computed the holographic entanglement entropy of $d+1$ dimensional rotating cylindrical black holes as perturbation over $AdS_{d+1}$. The significance of our paper is that the holographic entanglement entropy and the holographic entanglement first law of the rotating cylindrical black holes (for $d>2$) has not been calculated before. We derived the expression, based on variation of the minimal area functional, for the holographic entanglement entropy of any asymptotically AdS spacetime up to the first order of the metric perturbation. In the case of $d+1$ dimensional rotating cylindrical black holes, there is an interesting result in the entanglement entropy where the angular momentum appears at the first order of perturbation. Hence, the holographic first law was written with the presence of both energy and angular momentum to obtain the entanglement temperature and the entanglement angular velocities. In addition to the subsystem size $l$, the entanglement temperature is also found to be a function of rotation parameters (for $d>2$). The entanglement temperature was plotted as a function of the rotation parameter $a_1$ exhibiting a decreasing trend with increasing value of this parameter. Also, the entanglement angular velocities were computed as functions of rotation parameters. The entanglement angular velocities $\Omega_{1E}$ and $\Omega_{2E}$ were plotted as functions of the rotation parameters $a_1$ and $a_2$, respectively. In both cases, the entanglement angular velocity was found to increase with increasing rotation parameter.

\section*{Acknowledgements:}
We are grateful of Tadashi Takayanagi and Mohsen Alishahiha for careful reading of the draft and useful comments on the manuscript. We also acknowledge the scietific atmosphere of ICTP during the spring school of superstring and related topics 2019 where the last part of this work was done. 

%%%%%%%%%%%%%%%%%%%%%%%%%%%%%%%%%%%%%%%%%%%%%%%%%%%%%%%%%%%%%%%%%%%%%%%%%%%%%%%%%%

%%%%%%%%%%%%%%%%%%%%%%%%%%
%%%%%%%%%%%%%%%%%%%%%%%%%%%%%%%%%%%%%%%%%%%%%%%%%%%%%%%%%%%%%%%%%%%%%%%%%%%%%%%%%%
\section{Appendix}

\subsection{•}
To derive the geodesic equation, we start with the Eq. (\ref{deltaa}) where,
\bea\label{deltah}
\delta \sqrt{h}=\frac{1}{2} \sqrt{h} \ h^{ab} \delta h_{ab}.
\eea
The variation of the induced metric from Eq. (\ref{h}) with respect to the metric in the bulk and the coordinates may be recast as follows:
\bea\label{deltahab}
\delta h_{ab}&=&\big[\delta g_{\mu \nu}^{(0)} \partial_a x^{\mu} \partial_b x^{\nu}+g_{\mu \nu}^{(0)} \partial_a (\delta x^{\mu}) \partial_b x^{\nu}+g_{\mu \nu}^{(0)} \partial_a x^{\mu} \partial_b(\delta x^{\nu})\big].
\eea
By replacing Eqs. (\ref{deltah}) and (\ref{deltahab}) into Eq. (\ref{deltaa}), we obtain $\delta A$ in the following form:
\bea
\delta A&=&\frac{1}{2} \int d^{d-1} \zeta \ \sqrt{h}\ h^{ab} \big[\delta g_{\mu \nu}^{(0)} \partial_a x^{\mu} \partial_b x^{\nu}+g_{\mu \nu}^{(0)} \partial_a (\delta x^{\mu}) \partial_b x^{\nu}+g_{\mu \nu}^{(0)} \partial_a x^{\mu} \partial_b(\delta x^{\nu})\big].
\eea
Using $\delta g_{\mu \nu}^{(0)}=\frac{\partial g_{\mu \nu}^{(0)}}{\partial x^{\rho}}\delta x^{\rho}$, we get
\bea
\delta A&=&\frac{1}{2} \int d^{d-1} \zeta \big[\sqrt{h}\ h^{ab}\frac{\partial g_{\mu \nu}^{(0)} }{\partial x^\rho} \delta x^\rho \partial_a x^{\mu} \partial_b x^{\nu}+\partial_a(\sqrt{h}\ h^{ab} \ g_{\mu \nu}^{(0)} \delta x^\mu \partial_b x^{\nu})- \partial_a(\sqrt{h}\ h^{ab} \partial_b x^{\nu}) \ g_{\mu \nu}^{(0)} \delta x^\mu \\\nonumber
&-&\sqrt{h}\ h^{ab} \frac{\partial g_{\mu \nu}^{(0)}}{\partial x^\rho} \partial_a x^{\rho} \partial_b x^{\nu} \delta x^{\mu}+\partial_b(\sqrt{h}\ h^{ab} g_{\mu \nu}^{(0)}\partial_a x^{\mu} \delta x^\nu)-\partial_b(\sqrt{h}\ h^{ab} \partial_a x^{\mu}) g_{\mu \nu}^{(0)} \delta x^\nu\\\nonumber
&-&\sqrt{h}\ h^{ab} \frac{\partial g_{\mu \nu}^{(0)}}{\partial x^\rho} \partial_b x^{\rho} \partial_a x^{\mu} \delta x^{\nu}\big].
\eea
Ignoring the total derivative terms and simplifying somewhat slightly, we see that the condition of minimal surface, $\delta A=0$, leads to the Gauss equation
\bea\label{geodesic}
\partial_a(\sqrt{h}\ h^{a b} \partial_b x^{\mu}) +\sqrt{h}\ h^{ab} \Gamma^{\mu}_{\nu \rho} \partial_a x^{\nu} \partial_b x^{\rho} =0.
\eea
\subsection{ }
Here we intend to show that our approach  is equivalent to the covariant method proposed by Hubeny, Rangamani and Takayanagi (HRT) \cite{HRT}. The HEE of the 2+1 dimensional rotating BTZ black hole was obtained exactly by HRT in \cite{HRT}: 
\bea
S_{BTZ}=\frac{c}{6} \ln \Big(\frac{\beta_+ \beta_-}{\pi^2 \epsilon ^2} \sinh{(\frac{\pi l}{\beta_+})} \sinh{(\frac{\pi l}{\beta_-})}\Big),
\eea
where,  $\beta_{\pm}=\frac{2 \pi}{r_+ \pm r_-}$ and  $r_{+}, r_{-}$ are the outer and inner horizon respectively. Also $\epsilon$ is the UV cutoff and $c=\frac{3}{2 G}$ is the central charge of boundary dual $CFT_{2}$. So increasing the HEE of the subsystem with size $l$ in the presence of rotating BTZ black hole is obtained by subtracting above expression from pure AdS contribution, $S_{AdS_3}=\frac{c}{3} \ln \Big( \frac{l}{\epsilon} \Big)$:
\bea
\delta S=S_{BTZ}-S_{AdS_3}=\frac{c}{6} \ln \Big(\frac{\beta_+ \beta_-}{\pi^2 \l ^2} \sinh{(\frac{\pi l}{\beta_+})} \sinh{(\frac{\pi l}{\beta_-})}\Big).
\eea
 If we expand the above expression in the limit $\frac{\pi l}{\beta_{\pm}}<<1$, we get the following relation up to the first order of perturbation
\bea \label{m}
\delta S_1=\frac{l^2 (r_{+}^2+ r_{-}^2)}{48 G}.
\eea
On the other hand, if we use the formula (\ref{deltaAf}) for the 2+1 dimensional rotating BTZ, we see that the result is exactly matches with Eq. (\ref{m}). This was investigated in ref. \cite{Ghosh}.

Here we will show that the  HRT method is equivalent to our perturbative calculation of HEE in the higher dimensions. We use the covariant HRT method perturbatively and specifically for the rotating cylindrical black holes. For example in $d+1=4$ the metric (\ref{fff}) up to the first order of perturbation is as follows:
\bea
ds^2&=&\frac{dz^2}{z^2}-\frac{1}{z^2}\Big(1- \frac{2}{3} r_+^3 z^3 \Big) \Big(\Xi dt- a_1 d\tilde{x_1}\Big)^2\\\nonumber
&+&\frac{1}{z^2} \Big(1+\frac{1}{3} r_+^3 z^3\Big) \Big(a_1 dt -\Xi d\tilde{x_1}\Big)^2+\frac{1}{z^2} \Big(1+\frac{1}{3} r_+^3 z^3\Big) dx_1^2.
\eea
By assuming that $\tilde{x_1}$ and $t$ are the functions of $z$: $\tilde{x_1}(z), t(z)$, we can write the area functional in the following form
\bea \label{AP}
A=2\int_{\frac{-L}{2}}^{\frac{L}{2}} dx_1 \int_{0}^{z_{\star}} \sqrt{\Big(\frac{1}{z^2}+\frac{1}{3}r_+^3 z\Big)\Big(\frac{1}{z^2}+(\frac{-1}{z^2}+r_+^3 z(\Xi^2-\frac{1}{3}))t'^2-a_1 \Xi r_+^3 z  (t' \tilde{x_1}')+(\frac{1}{z^2}+r_+^3 z (a_1^2+\frac{1}{3}))\tilde{x_1}'^2\Big)}dz,
\eea
where, the prime denotes the derivative with respect to $z$. To minimize this area perturbtively, we start with pure AdS where $r_+^3=a_1=0$. In this case we can assume constant time slice ($t'=0$) for the minimal surface and write the above area functional as follows
\bea \label{Hamiltoni}
A= L \int_{\frac{-l}{2}}^{\frac{l}{2}} \frac{1}{z^2}\sqrt{1+z'^2}d\tilde{x_1}.
\eea
Considering the integrand of (\ref{Hamiltoni}) and $\tilde{x_1}$ as a Lagrangian and time, respectively, leads to a Hamiltonian:
\bea \label{Hamiltonii}
H=\frac{-1}{z^2 \sqrt{1+z'^2}}\  \ , \ \ \frac{dH}{d\tilde{x_1}}=0.
\eea
If we evaluate (\ref{Hamiltonii}) at $\tilde{x_1}=0$ where $z=z_{\star}$ and impose the condition  $z'(\tilde{x_1})=0$ at $z_{\star}$, we get to
\bea \label{x,t}
\tilde{x_1}'(z)=\frac{d\tilde{x_1}}{dz}=\pm \frac{z^2}{\sqrt{z_{\star}^4-z^4}}\  \ , \ \ t'=0.
\eea
By integrating $d\tilde{x_1}$ from $\frac{-l}{2}$ to $\frac{l}{2}$, we can write $z_{\star}$ in terms of $l$: $z_{\star}=\frac{l\ \Gamma(\frac{1}{4})}{2 \sqrt{\pi} \Gamma(\frac{3}{4})}$ . Substituting Eqs. (\ref{x,t}) to  Eq. (\ref{AP}) and expanding up to the first order of $r_{+}^3$ gives
\bea
\delta S_1=\frac{\delta A}{4 G_N^4}=\frac{ r_+^3 l^2 L(1+a_1^2) \Gamma(\frac{1}{4})^2}{128 G_N^4 \Gamma(\frac{3}{4})^2},
\eea
where $\delta$ denotes the difference between first order of perturbation and pure AdS contribution. The above relation exactly matches with Eq. (\ref{entropy2}) for $d=3$ and $n=1$. Also for higher dimensions, compatibility can be shown in this way. Note that if we continue this perturbative method to the second order, we need to write the metric expansion, Eq. (\ref{fff}) up to the second order of perturbation, and more importantly Eqs. (\ref{x,t}) for $\tilde{x_1}'(z)$ and $t'$ must be corrected by terms that contain the first order of perturbation.
\subsection{}
Consider $f$, $g$, $h$, and $k$ as functions of $(q_1,q_2,q_3)$. Now, we can use the following  relation for the derivative of $f$ with respect to $g$ while $h$ and $k$ are fixed:
\bea
\Big(\frac{\partial f}{\partial g}\Big)_{h,k}=\frac{\lbrace f,h,k\rbrace_{q_1,q_2,q_3}}{\lbrace g,h,k\rbrace_{q_1,q_2,q_3}},
\eea
where, the Nambu bracket is defined as follows:
\bea
\lbrace f,h,k\rbrace_{q_1,q_2,q_3}=\sum_{ijk=1}^{3} \epsilon_{ijk} \frac{\partial f}{\partial q_i}\frac{\partial h}{\partial q_j}\frac{\partial k}{\partial q_k}=\begin{vmatrix}
\frac{\partial f}{\partial q_1}&\frac{\partial f}{\partial q_2}&\frac{\partial f}{\partial q_3}\\
\frac{\partial h}{\partial q_1}&\frac{\partial h}{\partial q_2}&\frac{\partial h}{\partial q_3}\\
\frac{\partial k}{\partial q_1}&\frac{\partial k}{\partial q_2}&\frac{\partial k}{\partial q_3}
\end{vmatrix}.
\eea
When $f$, $g$, and $h_n (n=1,2,3,...)$ are functions of $(n + 1)$ variables, the generalization of the above relations becomes \cite{Hosseini}:
\\
\bea
\Big(\frac{\partial f}{\partial g}\Big)_{h_1,....,h_n}=\frac{\lbrace f,h_1,...,h_n\rbrace_{q_1,q_2,....,q_{n+1}}}{\lbrace g,h_1,...,h_n\rbrace_{q_1,q_2,....,q_{n+1}}},
\eea
where,
\bea
\lbrace f,h_1,...,h_n\rbrace_{q_1,q_2,....,q_{n+1}}=\sum_{ijk....l=1}^{n+1} \epsilon_{ijk...l} \frac{\partial f}{\partial q_i}\frac{\partial h_1}{\partial q_j}\frac{\partial h_2}{\partial q_k}.......\frac{\partial h_n}{\partial q_l}.
\eea

%%%%%%%%%%%%%%%%%%%%%%%%%%%%%%%%%%%%%%%%%%

\end{document}